\documentclass{emulateapj}
\usepackage{graphicx}
\usepackage{pdfpages}
\usepackage{tablefootnote}
\usepackage{epstopdf}
\usepackage{natbib}
\usepackage{amsmath}
\newcommand{\mstar}{$M_*~$}
\newcommand{\mhalo}{$M_{halo}$}
\newcommand{\mstareq}{M_*}
\newcommand{\mhaloeq}{M_{halo}}

\newcommand{\msuneq}{M_{\odot}}
\newcommand{\rvir}{$r_{vir}$}
\newcommand{\rvireq}{r_{vir}}

\newcommand{\mratio}{\left( \frac{\mhaloeq}{M_1} \right)}
\newcommand{\beq}{\begin{equation}}
\newcommand{\eeq}{\end{equation}}

\begin{document}

\title{A Deep Search for Faint Galaxies Associated with Very Low-Redshift C IV Absorbers: A Case with Cold-Accretion Characteristics\altaffilmark{*}}
\author{Joseph N. Burchett\altaffilmark{1}$^,$\altaffilmark{2}, Todd M. Tripp\altaffilmark{2}, Jessica K. Werk\altaffilmark{3}, J. Christopher Howk\altaffilmark{4}, J. Xavier Prochaska\altaffilmark{3}, Amanda Brady Ford\altaffilmark{5}, and Romeel Dav\'e\altaffilmark{6}$^,$\altaffilmark{7}$^,$\altaffilmark{8}}

\altaffiltext{1}{Email: jburchet@astro.umass.edu}
\altaffiltext{2}{Department of Astronomy, University of Massachusetts, 710 North Pleasant Street, Amherst, MA 01003-9305}
\altaffiltext{3}{UCO/Lick Observatory, University of California, Santa Cruz, CA}
\altaffiltext{4}{Department of Physics, University of Notre Dame, Notre Dame, IN 46556}
\altaffiltext{5}{Astronomy Department, University of Arizona, Tucson, AZ 85721, USA}
\altaffiltext{6}{University of the Western Cape, Bellville, Cape Town 7535, South Africa}
\altaffiltext{7}{South African Astronomical Observatories, Observatory, Cape Town 7925, South Africa}
\altaffiltext{8}{African Institute for Mathematical Sciences, Muizenberg, Cape Town 7945, South Africa}
\altaffiltext{*}{Based on observations obtained with the NASA/ESA Hubble Space Telescope operated at the Space Telescope Science Institute, which is operated by the Association of Universities for Research in Astronomy, Inc., under NASA contract NAS5-26555. Also, based on data acquired using the Large Binocular Telescope (LBT). The LBT is an international collaboration among institutions in the US, Italy, and Germany. LBT Corporation partners are the University of Arizona, on behalf of the Arizona University System; Instituto Nazionale do Astrofisica, Italy; LBT Beteiligungsgesellschaft, Germany, representing the Max Planck Society, the Astrophysical Institute of Potsdam, and Heidelberg University; Ohio State University, and the Research Corporation, on behalf of the University of Notre Dame, the University of Minnesota, and the University of Virginia. Observations reported here were obtained at the MMT Observatory, a joint facility of the University of Arizona and the Smithsonian Institution.}

\begin{abstract}
Studies of QSO absorber-galaxy connections are often hindered by inadequate information on whether faint/dwarf galaxies are located near the QSO sight lines. To investigate the contribution of faint galaxies to QSO absorber populations, we are conducting a deep galaxy redshift survey near low$-z$ C IV absorbers. Here we report a blindly-detected \ion{C}{4} absorption system ($z_{\rm abs} = 0.00348$) in the spectrum of PG1148+549 that appears to be associated either with an edge-on dwarf galaxy with an obvious disk (UGC 6894, $z_{\rm gal}$ = 0.00283) at an impact parameter of $\rho$ = 190 kpc or with a very faint dwarf irregular galaxy at $\rho$ = 23 kpc, which is closer to the sightline but has a larger redshift difference ($z_{\rm gal} = 0.00107$, i.e., ~$\delta v$ =724 km/s). We consider various gas/galaxy associations, including infall and outflows.  Based on current theoretical models, we conclude that the absorber is most likely tracing (1) the remnants of an outflow from a previous epoch, a so-called ‘ancient outflow’, or (2) intergalactic gas accreting onto UGC 6894, ‘cold mode’ accretion. The latter scenario is supported by \ion{H}{1} synthesis imaging data that shows the rotation curve of the disk being codirectional with the velocity offset between UGC6894 and the absorber, which is located almost directly along the major axis of the edge-on disk.
\end{abstract}

\section{Introduction}
The interactions of galaxies with their ambient surrounding media and with one another have come into sharp focus as crucial components of galaxy evolution. These interactions include the continuing accretion of material required to fuel on-going star formation and, conversely, the feedback mechanisms that regulate galactic physical conditions and transport metal-enriched gas to galactic halos/circumgalactic media (CGM) and beyond \citep[e.g.,][]{Fumagalli:2011qy,  Hopkins:2006yq, Veilleux:2005lr, Heckman:2011qy}.  Although inflow and outflow processes are challenging to observe, QSO absorption spectroscopy provides a sensitive tool for doing so. 

However, discerning the origin of the gas detected in QSO absorption spectra is greatly complicated by the incompleteness of galaxy redshift surveys in the fields of the absorbers.  For instance, the Sloan Digital Sky Survey's 95\% spectroscopic completeness down to $m_r=17.7$ \citep{Strauss:2002fk} includes only $L>L_*$ galaxies at z$\geq$0.15.  Thus, while the local Universe provides the most suitable laboratory for studying the galaxy environments of intervening gaseous systems, the possibility remains of attributing the detected gas to a more luminous galaxy when a fainter, undetected dwarf galaxy is present.  These concerns have important implications because while an absorber might appear to arise in some type of inflow or outflow connected to a luminous galaxy, it could in fact be bound to a faint satellite or nearby dwarf galaxy that was overlooked in the incomplete redshift survey.

\begin{deluxetable*}{cccccc} 
\tabletypesize{\scriptsize} 
\tablewidth{0pt} 
\tablecaption{Absorption line properties} 
\tablehead{\colhead{Ion} & \colhead{$\lambda_{0}$ (\AA)} & \colhead{$W_r$ (m\AA)\tablenotemark{a}}  & \colhead{$\text{log}~N$ (cm$^{-2})$\tablenotemark{b}} & \colhead{$b$ (km/s)} & \colhead{$v$ (km/s)\tablenotemark{c}}} 
\startdata 
H \sc{i} & 1215.67 & 375.7 $\pm$ 5.55 & $>$14.41 & 30  $\pm$ 1 &  0  \\  
H \sc{i} & 1215.67 & 134.51 $\pm$ 5.52 & 13.51 $\pm$ 0.02 & 29  $\pm$ 1 & 134 \\  
C \sc{iv} & 1548.2 & 89.84 $\pm$ 7.19 & 13.64 $\pm$ 0.03 & 9  $\pm$ 1 & -9 \\  
C \sc{iv} & 1550.77 & 58.39 $\pm$ 7.27 & 13.64 $\pm$ 0.03 & 9  $\pm$ 1 & -9 \\  
C \sc{ii} & 1334.53 & $<$71.14 & $<$13.59 &  -  &  -  \\  
Si \sc{iv} & 1393.75 & $<$68.45 & $<$12.92 &  -  &  -  \\  
Si \sc{iv} & 1402.77 & $<$86.12 & $<$13.33 &  -  &  -  \\  
Si \sc{iii}\tablenotemark{d} & 1206.5 &  -  &  -  &  -  &  -  \\  
Si \sc{ii} & 1260.42 & $<$53.94 & $<$12.54 &  -  &  -  \\  
\enddata 
\tablenotetext{a}{Nondetections are reported as 3$\sigma$ upper limits.} 
\tablenotetext{b}{Column density upper limits are calculated from the $W_r$ limits assuming a linear curve-of-growth relationship.} 
\tablenotetext{c}{Velocity offsets are measured relative to the strong Ly$\alpha$ feature.} 
\tablenotetext{d}{We report a nondetection of Si \textsc{III} because, if present, this line is blended with an \textsc{O VI} feature at another redshift.} 
\end{deluxetable*}

At very low redshifts, the \ion{C}{4} $\lambda \lambda~1548.2, 1550.8$ doublet provides an easily identifiable signature of metal-enriched gas in QSO spectra.  Space-based observations are required at these wavelengths in the nearby universe, and the Cosmic Origin Spectrograph (COS) aboard the Hubble Space Telescope \citep{Green:2012qy} is  the most sensitive space-based instrument to-date for this work.  As part of a larger, blind survey of \ion{C}{4} absorbers at low redshift, we have discovered a z=0.003 \ion{C}{4} absorption system whose location and/or kinematics, in light of previous observations, suggest two interesting galaxy associations: the absorber is (1) at a similar redshift but at 1.4 virial radii from a normal star-forming galaxy or (2) at an impact parameter of 23 kpc from a faint dwarf galaxy but with a velocity separation of $\delta v \sim 700 ~\text{km/s}$.  In this letter, we report our analysis of this absorber, and throughout we assume a cosmology of $H_0 = 72 ~\text{km/s~Mpc}^{-1}$, $\Omega_M=0.27$, and $\Omega_{\Lambda}=0.73$.

\section{Observations}
The absorber of interest is a \ion{C}{4} absorption system detected in high signal-to-noise (S/N) HST/COS spectra, and the corresponding galaxy survey employs publicly available data from the SDSS, imaging with the Large Binocular Telescope (LBT), and spectroscopy with the MMT+Hectospec \citep{Fabricant:1998uq}, Keck+DEIMOS \citep{Faber:2003kx} and Shane 3m+Kast \citep{KastManual}.
	
\subsection{HST/COS Spectroscopy}
The system appears at redshift $z_{\rm abs} = 0.00348$ in the spectrum of PG1148+549 ($z_{\rm QSO} = 0.9754$), which was observed as part of the HST program 11741 using the Cosmic Origins Spectrograph. The observations and data reduction are described by \cite{Meiring:2013fj,Meiring:2011fj}.  At the absorber redshift, the spectrum has signal-to-noise per resolution element S/N = 30 at the wavelength of the Ly$\alpha$ line and S/N = 28 near the \ion{C}{4} doublet, which enable us to detect \ion{C}{4} lines with rest-frame equivalent width $W_r$ $\gtrsim$ 10 m\AA.  The Ly$\alpha$ and \ion{C}{4} lines detected in the COS spectrum at $z_{\rm abs} = 0.00348$ are shown in Figure 1.

\subsection{Optical Imaging and Spectroscopy}
To seek associated faint or low-surface brightness galaxies with greater depth than SDSS, we obtained deep broadband imaging of the field using the Large Binocular Telescope (LBT) with a limiting magnitude of B$_{AB}$ $\sim$ 25.5 \citep{Meiring:2013fj}.

We also obtained follow-up multi-object spectroscopy using Hectospec to measure fainter galaxy redshifts.  Two low-surface brightness objects were visible in the LBT imaging but were not measured with Hectospec, and we obtained their slit spectra from Keck/DEIMOS and the Kast Double Spectrograph on the Lick 3.0-m telescope.

\section{Analysis}
The UV spectrum of PG1148+549 was visually inspected for absorption systems by two members of our team independently, and we unambiguously identify the \ion{C}{4} doublet and Ly$\alpha$ absorption at z=0.00348. \ion{Si}{3} absorption cannot be measured due to blending. We fit Voigt profiles to the lines to determine column densities, Doppler parameters, and velocity centroids; the \ion{C}{4} doublet and Ly-$\alpha$ line are shown in Figure 1 along with the superimposed apparent column density profiles \citep{Savage:1991vn} of the doublet, and the corresponding measurements are listed in Table 1.  Although Ly$\alpha$ appears to be unsaturated in Figure 1, we report the \ion{H}{1} column density measurements as lower limits because the line spread function of COS is known to have large wings that can fill in the cores of deep lines \citep{COSLSF}.  Because we do not have wavelength coverage for the higher Lyman series lines, we opt for this conservative estimate.  With only a lower limit on the \ion{H}{1} column density and a precise measurement of a single metal (\ion{C}{4}), we cannot determine the absorber metallicity.  Even if we constrain $n_H$ [based on our limit on N(\ion{C}{4})/N(\ion{C}{2})] and then use $n_H$ to estimate N(\ion{H}{1}) by assuming hydrostatic equilibrium \citep{Schaye:2001fj}, we still find that a very large range of metallicities is allowed.

\begin{figure*}[ht]
\centering
\includegraphics[width=0.8\textwidth]{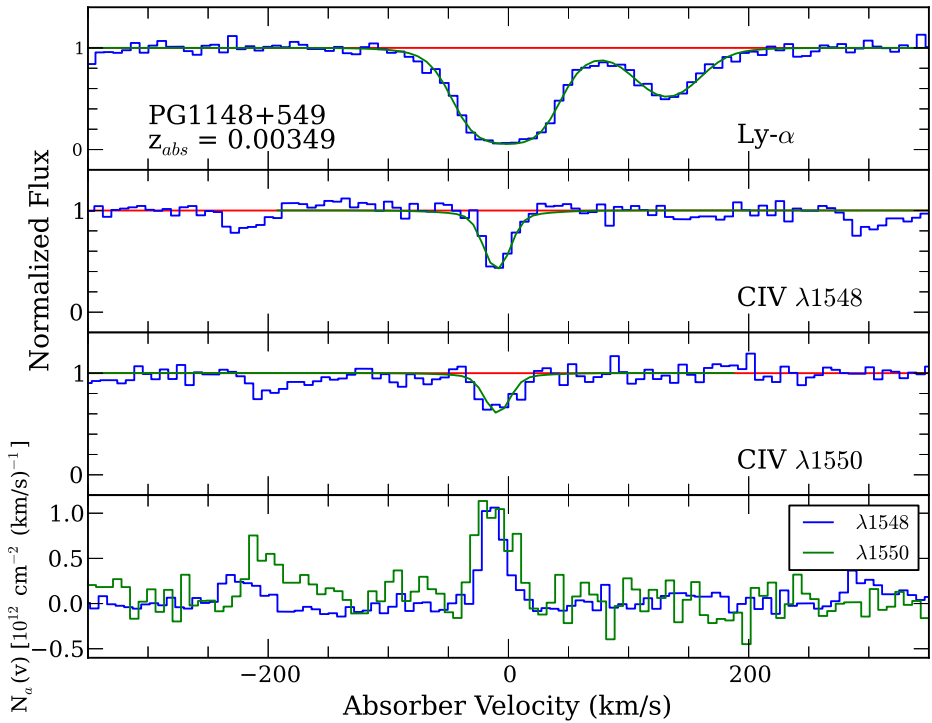}
\caption{Normalized spectrum showing the \ion{C}{4} doublet along with the Ly$\alpha$ line for the z=0.00348 absorber.  The bottom pane shows the superimposed apparent column density profiles of the \ion{C}{4} $\lambda$ 1548 (blue) and $\lambda$ 1550 (green) lines. }
\label{fig:spec}
\end{figure*}

To search for associated galaxies, we used data from the SDSS along with the follow-up observations described in Section 2.  For galaxies brighter than $m_b = 18.2$, our survey is 86\% complete out to impact parameter $\rho=180$ kpc (corresponding to the 1$^\circ$ field of view of Hectospec at the absorber redshift). This magnitude limit corresponds to L = 0.0008 L$_*$ at the absorber redshift.  Table 2 lists the galaxies with measured redshifts within 400 kpc of the absorber with $|\delta v|\leq900$ km/s of the \ion{C}{4} doublet heliocentric velocity.  \citet{Verheijen:2001lr} report a mean distance of 18.6 Mpc for UGC 6894 and galaxies in this vicinity; therefore, the galaxy impact parameters and luminosity (as a fraction of $L_*$) corresponding to this distance are listed, with the exception of LBT J115205.6+544732.2, which was calculated at the Hubble flow luminosity distance corrected for the Local Group (LG) velocity, Virgo infall, the Great Attractor, and the Shapley Supercluster \citep{Mould:2000vn}.

Figure 2 shows the galaxies with measured redshifts within 400 kpc of the absorber and with $|\delta v| \leq  900$ km/s.  Here, we call attention to the two closest galaxies to the sightline: UGC 6894 lying almost directly to the east of the sightline and LBT J115205.6+544732.2, which we refer to as ``the Grapes" hereafter, lying just northeast of the sightline.  UGC 6894 is separated in velocity by -196.9 $\pm$ 15.1 km/s and by 190 kpc in impact parameter; the Grapes is only 23 kpc away but has a velocity difference of -724 $\pm$ 210 km/s.  We consider these two galaxies the most likely candidates for association with the absorber because there are no other galaxies brighter than 0.001 L$_*$ closer in impact parameter (kpc).  However, we note that a group of luminous and actively star-forming galaxies is found at somewhat larger projected distances northeast of the QSO (see Figure 2),  two of which are at similar impact parameter in virial radius units as UGC 6894 but are nearly 100 kpc further from the sightline. 

We wish to place these potential galaxy/absorber associations into a physical context related to the properties of the galaxies themselves, including their stellar masses (\mstar) and halo masses (\mhalo). Therefore, we employ the stellar/halo mass relation of \cite{Moster:2013lr} combined with a stellar mass calculation from the \textsc{kcorrect} software \citep[estimated errors for low mass galaxies: $\pm 50 \%$]{Blanton:2007ys}, which fits stellar population synthesis models to broadband photometry (we use SDSS+2MASS photometry). \textsc{kcorrect} assumes a Chabrier IMF and $H_0 = 100  ~\text{km/s Mpc}^{-1}$, and the \citet{Moster:2013lr} formalism also assumes a Chabrier IMF but with $H_0 = 70.4 ~\text{km/s Mpc}^{-1}$.  We use the stellar mass calculation of \citet{McIntosh:2008fk}, which simply uses broadband colors, to calculate the ratio of the masses when the absolute magnitudes are scaled using these differing cosmologies.  Finally, we scale the stellar mass from \textsc{kcorrect} and solve the stellar mass/halo mass ratio model derived by \cite{Moster:2013lr} for the halo mass:

\begin{equation}
\frac{\mstareq}{\mhaloeq} = 2 \left( \frac{m}{M} \right)_0 \left[ \mratio^{-\beta} + \mratio^\gamma \right]^{-1}
\end{equation}
where we use their fitted values of log $M_1$ = $11.594$, $(m/M)_0$ = $0.0350$, $\beta$ = $1.3735$, and $\gamma$ = 0.6090, corresponding to z=0.003.  The virial radius, defined by a factor of 200 overdensity, is then
\beq
\rvireq= \left( \frac{\mhaloeq}{200~(4\pi/3)~\rho_\text{crit}} \right)^\frac{1}{3}
\eeq
Using this formulation, we calculate $log~ M_* = 8.62~\msuneq$, $log~\mhaloeq = 10.83~\msuneq$, and \rvir = 133 kpc for UGC 6894.  For comparison, we also calculated the virial radius using three other methods \citep{Prochaska:2011yq,Werk:2012qy,Stocke:2013mz} to find \rvir = 116 kpc, 160 kpc, and 70 kpc.

\begin{figure*}[!t]
\centering
\includegraphics[width=1.0\textwidth]{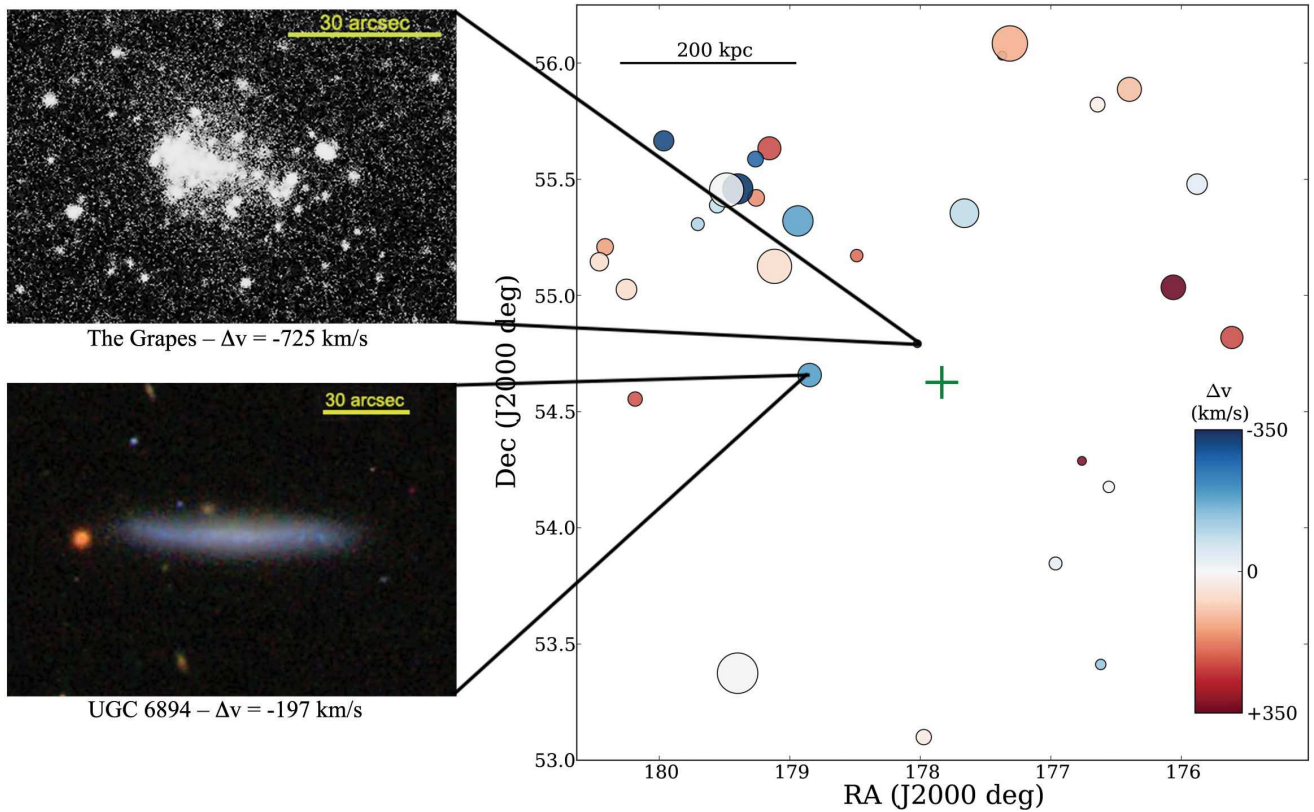}
\caption{The field of PG1148+547 showing all galaxies with measured redshifts within 400 kpc impact parameter to the sightline and 900 km/s of the absorber.  The marker colors indicate the recession velocity difference between the galaxy and absorber, and marker sizes indicate the luminosity of the galaxy.  For reference, the lower-leftmost galaxy (M109) plotted has luminosity L $\sim$ 2 L*.   The two galaxies primarily discussed in this analysis are the Grapes (top image on the left) and UGC 6894 (bottom image on the left).}
\label{fig:spec}
\end{figure*}

\section{Discussion}

\subsection{The Grapes}
The Grapes is a dwarf irregular galaxy located at z=0.00107 $\pm$ 0.0007 with an impact parameter of 23 kpc.  The redshift was measured from the H$\alpha$ emission line found in the Lick/Kast spectrum, and we measure a minimum star formation rate (SFR) of $5.9 \times 10^{-6}~\msuneq/yr$ from this line \citep{Kennicutt:1998fj}.  This is a minimum SFR for two reasons: first, we do not measure the H$\beta$ emission and therefore cannot correct for dust reddening; second, the Grapes is very clumpy, and our long-slit spectrum did not cover all star clusters in the galaxy.  SDSS (DR10) detected the Grapes in its imaging, but resolved it into three individual clumps, two of which are labeled as faint galaxies.  From the SDSS photometry, we estimate stellar masses of \mstar = $4.8\times10^5\msuneq$, $9.1\times10^3\msuneq$, and $2.3\times10^4\msuneq$ for these three clumps.  This third clump, which SDSS classified as an interloping star, differs in \emph{u-g} color from the other two, which could indicate that this clump is not associated with the galaxy; also, it is a much fainter object with large photometric uncertainties.

While the Grapes has by far the smallest projected distance and therefore is a prime candidate for the absorbing gas source, the galaxy is separated in velocity by $\delta v=724\pm210~\text{km/s}$, where uncertainties of 15 km/s and 210 km/s arise from the COS wavelength calibration and the Kast H$\alpha$ measurement, respectively.  This large $\delta v$ would seem to suggest that the absorber and galaxy are unrelated, but both models and observations indicate that galaxies can drive outflows with velocities of hundreds of km/s \citep{Tremonti:2007fk,Murray:2011lr,Rupke:2013yq,Rubin:2013lr}.  Of course, the velocity observed is only the radial component, and this $\delta v$ is a lower limit to the total velocity offset.  For a plausible association, one must invoke some sort of feedback process, such as a galactic superwind in which the gas might be entrained.  We implemented the formalism of \cite{Murray:2010pd,Murray:2011lr} to test the plausibility of this galaxy driving a $> 500$ km/s wind.  Even while allowing strong contributions from protostellar jets, overestimating the typical cluster mass (using the mass of the largest clump) and supernova rate, and using a likely gross overestimate of the total galaxy luminosity, we find that we must neglect any resistance from the galaxy ISM, such as turbulence, to accelerate a wind to 200 km/s at comparable impact parameters.  On this basis, we conclude that it is unlikely that the Grapes could drive an outflow to sufficient speed to explain the large redshift difference between the absorber and the galaxy.

\subsection{UGC 6894}
UGC 6894 is located on the very outer perimeter of the Ursa Major Cluster \citep{Tully:1987uq}, which raises some concerns. For example, the halo mass/virial radius calculation could be invalid if the galaxy is actually within the much larger halo of a cluster.  However, \citet{Tully:1987uq} define this cluster based purely on spatial overdensity and the members' similar recession velocities, not because the members possess the typical cluster environment characteristics in morphology and velocity dispersion. This region of the sky is dominated by gas-rich, late-type spiral galaxies unlike true clusters, which are dominated by early-type members.  Furthermore, using the quoted cluster virial radius from \citet{Tully:1987uq}, UGC 6894 is located $>2~r_{vir}$ from the apparent spatial center.

\begin{figure*}[!t]
\centering
\includegraphics[width=1.0\textwidth]{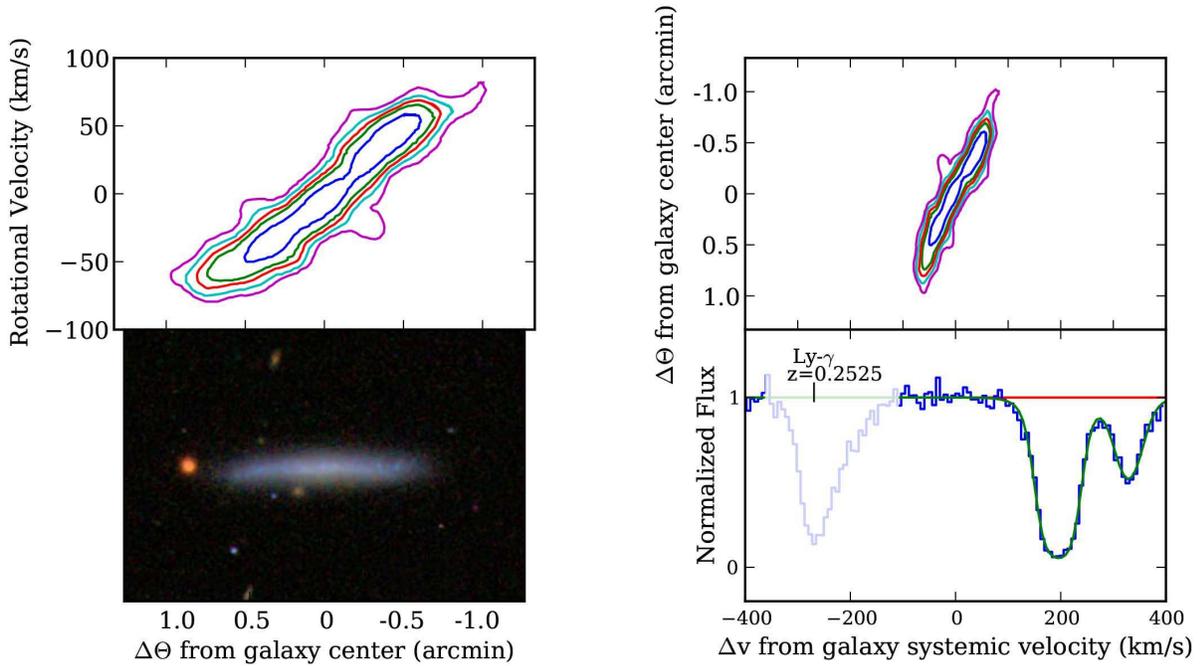}
\caption{(Left) A color composite image of UGC 6894 from the SDSS along with the \ion{H}{1} rotation curve from the WSRT.  The QSO sightline is off the panel to the right.  The galaxy's rotation in this spatial direction is receding (into the plane of the sky), in the same direction as the absorber/galaxy velocity offset. Note that the detected \ion{H}{1} emission extends only slightly beyond the optical disk.  (Right) The rotation curve from \ion{H}{1} line emission along with the absorber Ly$\alpha$ line in the frame of the galaxy systemic velocity.  The galaxy's gaseous rotation curve rises through the last point measured, at 63 $\pm$ 5 km/s, and the velocity offset of the absorber is in the same direction as the galaxy's rotation. One arcmin corresponds to $\sim$5.4 kpc, and the outer contours on the \ion{H}{1} maps correspond to $\sim2 \times 10^{20}$ atoms cm$^{-2}$. Position-velocity data adapted from \citet{Verheijen:2001lr}.}
\label{fig:spec}
\end{figure*}

We assume an 18.6 Mpc distance \citep{Verheijen:2001lr} for UGC 6894, and we note that this distance is much greater than that indicated by a pure Hubble flow from the heliocentric redshift (12.0 Mpc) but is similar to that obtained when correcting for the LG velocity, etc. (17.1 Mpc). At a distance of 18.6 Mpc, the corresponding angular scale is 5.41 kpc/arcmin, and the galaxy's 35.210 arcmin  angular distance from the sightline then corresponds to an impact parameter of 190.5 kpc.  The recession velocities of the absorber and galaxy differ by $196.9\pm15.1$ km/s, where the uncertainty is dominated by the COS wavelength calibration uncertainty of 15 km/s \citep{COShandbook}; \citet{Verheijen:2001lr} report a redshift uncertainty of 2 km/s for UGC 6894.  As shown in Figure 2, the QSO sightline passes almost directly to the east, lying along a direction oriented just 2$^{\circ}$ south of east from the edge-on major axis of UGC 6894.  
Based on the four virial radii calculated in Section 3, the absorber appears to be located 1.2-2.7 \rvir ~from the center of UGC 6894. 

By using Eqs. 1 \& 2 of \cite{Dave:2010qy}, we calculate the threshold virial overdensity relative to the critical density at the absorber redshift $\delta_{\text{th}} \sim 120$.  From our lower limit on N(\ion{H}{1}) and Eq. 3 of \cite{Dave:2010qy}, we find that \text{log} N(\ion{H}{1}) = 14.4 corresponds to an overdensity $\rho/\bar{\rho}\sim 94$, which is consistent with association at a distance just outside the virial radius of a galaxy halo.  If the absorber is associated with UGC 6894, one can consider three possible scenarios: (1) the absorber traces gas that is currently neither inflowing nor outflowing, e.g., residing in a gaseous halo, possibly a result of much earlier outflows, (2) the gas is being ejected from the galaxy, or (3) the absorber is tracing accretion from the IGM onto the galaxy.  

In considering scenario (1), we question whether $\sim$190 kpc is a reasonable extent for the gaseous disk. \cite{Sheth:2010ly} measure the diameter of the 25th magnitude isophote in the B-band ($D_{25}$) to be 1.66 arcmin, which corresponds to $\sim$9 kpc.  Therefore, 42x this isophotal radius is well beyond a reasonably expected extent of the disk.  This absorber/galaxy pair features a somewhat large extent for a \ion{C}{4} absorber compared to other studies (Bordoloi et al. 2013, in preparation; \citealp{Chen:2001lr, Chen:2009vn, Stocke:2013mz}). \cite{Tripp:2006qy} suggested that \ion{O}{6}/\ion{C}{4} systems they detected far away from galaxies may be metal-enriched gas from outflows that occurred in previous epochs.  This claim is supported by recent simulations from Ford et al. (2013, submitted).  For example, these simulations show a mean column density of N(\ion{C}{4})=$10^{13}~\text{cm}^{-2}$ at $\sim$190 kpc for gas attributed to `ancient outflows' in an $\mhaloeq=10^{11}\msuneq$ galaxy, similar to the halo mass of UGC 6894. This material could also be residual gas from dynamical processes such as tidal stripping.

Scenario (2) suffers from several problems.  First, the orientation of the galaxy is such that the sightline falls nearly 90$^{\circ}$ from the polar axis of the galaxy, where biconical outflows are most likely to propagate perpendicular to the disk \citep{Rubin:2013lr, Bordoloi:2011uq, Bouche:2012kx, Kacprzak:2012kx}.  Also, if the gas is outflowing coplanarly with the UGC 6894 disk and was expelled from the more stellar-dense visible disk, the gas would have had to travel directly through the relatively dense medium of the gaseous/stellar disk for billions of years at 100s of km/s all while gaining angular momentum.  However, it is possible that this \ion{C}{4} absorber is the detritus from a galactic-fountain flow or from tidal interactions that has settled into the disk and is now returning to the galaxy.

Scenario (3) provides perhaps the most compelling comparison with theory.  Figure 3 shows the \ion{H}{1} rotation curve of UGC 6894 obtained using aperture synthesis \citep{Verheijen:2001lr}, and the side of the disk closest to the absorber is receding (into the plane of the sky).  Consequently, the absorber velocity is roughly consistent with an extension of the rotation curve to a larger angular separation.  Such a configuration was predicted by \cite{Stewart:2011gf} as an observational signature of cool gas accreting into a galaxy via a long, warped disk stream that adds angular momentum and mass to the galaxy.  Absorber velocity separations consistent with the rotation of a disk galaxy have been observed at z=0.4-0.6 by \cite{Steidel:2002ve}, but at smaller impact parameters ($\rho \lesssim$75 kpc).  \cite{Stewart:2011gf} predict this accretion signature to be observed at $\rho \sim \rvireq/3$ at z=0 for $L_*$ galaxies because the transition to hot mode accretion is believed to have occurred \citep{Stewart:2011ul} by z=0, as enough mass will have accreted onto the galaxy to exceed the cold/hot mode threshold mass ($\mhaloeq \sim 10^{12}$).  Therefore, the gas in the outer halo will have been virialized and shock-heated, removing the cool gas signature from the regions near the virial radius.  
However, UGC 6894 is well below this threshold mass, and we suggest that the absorber could be cool gas accreting onto the galaxy; at low redshifts, cold accretion could occur predominantly in lower mass halos reflecting the shift of star-formation activity to lower mass galaxies at the present epoch (the so-called “downsizing” trend).  To probe this topic with a larger, statistically significant sample, we are now conducting a follow-up survey of the redshifts and properties of galaxies in the vicinity of low-z \ion{C}{4} absorbers blindly identified in COS spectra.  The larger survey results will be reported in a forthcoming paper once the follow-up observations are completed.
\\ 
\\ \\
We thank Neal Katz, John O'Meara, and the referee for helpful discussions and Scott Lange for assistance with the Keck observations. The authors wish to recognize and acknowledge the very significant cultural role and reverence that the summit of Mauna Kea has always had within the indigenous Hawaiian community. We are most fortunate to have the opportunity to conduct observations from this mountain. This research was supported by NASA grant HST-GO-11741 from the STScI and by NSF grant AST-0908334 and AST-1212012.

\begin{deluxetable*}{ccccccc} 
\centering
\tabletypesize{\scriptsize} 
\tablewidth{0pt} 
\tablecaption{Galaxies potentially associated with the absorber at $z_{abs}=0.00348$.} 
\tablehead{\colhead{Galaxy} & \colhead{$\alpha_{gal}$ (deg J2000)} & \colhead{$\delta_{gal}$ (deg J2000)} & \colhead{$z_{gal}$}  & \colhead{$\delta$v (km/s)} & \colhead{$\rho$ (kpc)\tablenotemark{a}} & \colhead{L/L$_*$\tablenotemark{a}}}  
\startdata 
UGC 06894 & 178.84787 & 54.65734 & 0.00283 & -197  & 190  & 0.02 \\  
SDSS J115356.95+551017.3 & 178.48733 & 55.17150 & 0.00417 & 202  & 214  & 0.002 \\  
SDSS J114702.85+541716.8 & 176.76188 & 54.28800 & 0.00458 & 326  & 231  & 0.0006 \\  
NGC 3913 & 177.66225 & 55.35386 & 0.00318 & -92  & 238  & 0.08 \\  
SDSS J114613.45+541034.3 & 176.55604 & 54.17619 & 0.00348 & -2  & 283  & 0.002 \\  
NGC 3982 & 179.11720 & 55.12524 & 0.00370 & 62  & 287  & 0.2 \\  
SDSS J114751.36+535047.9 & 176.96401 & 53.84666 & 0.00339 & -30  & 303  & 0.002 \\  
NGC 3972 & 178.93787 & 55.32074 & 0.00284 & -193  & 303  & 0.1 \\  
NGC 3846A & 176.06177 & 55.03497 & 0.00479 & 387  & 355  & 0.04 \\  
SDSS J115701.86+552511.2 & 179.25775 & 55.41981 & 0.00405 & 168  & 367  & 0.005 \\  
NGC 3990 & 179.39816 & 55.45867 & 0.00232 & -349  & 394  & 0.1 \\  
SDSS J115813.69+552316.5 & 179.55708 & 55.38794 & 0.00322 & -80  & 402  & 0.005 \\  
NGC 3998 & 179.48389 & 55.45359 & 0.00347 & -6  & 405  & 0.9 \\  
UGC 06919 & 179.15629 & 55.63319 & 0.00428 & 235  & 406  & 0.03 \\  
SDSS J115703.08+553512.3 & 179.26285 & 55.58677 & 0.00255 & -282  & 407  & 0.005 \\  
SDSS J115849.18+551824.8 & 179.70493 & 55.30689 & 0.00313 & -107  & 410  & 0.003 \\  
SBS 1139+550 & 175.61338 & 54.81901 & 0.00428 & 236  & 420  & 0.02 \\  
MCG +09-20-060 & 180.18479 & 54.55422 & 0.00424 & 225  & 442  & 0.003 \\  
SDSS J114634.06+554917.0 & 176.64195 & 55.82140 & 0.00357 & 22  & 444  & 0.003 \\  
UGC 06685 & 175.87975 & 55.47891 & 0.00334 & -43  & 453  & 0.01 \\  
SDSS J114628.27+532443.5 & 176.61783 & 53.41208 & 0.00303 & -136  & 459  & 0.002 \\  
SDSS J114929.68+560154.6 & 177.37367 & 56.03183 & 0.00306 & -127  & 464  & 0.001 \\  
MCG +09-20-063 & 180.25146 & 55.02583 & 0.00369 & 60  & 467  & 0.01 \\  
NGC 3898 & 177.31404 & 56.08436 & 0.00392 & 129  & 482  & 0.9 \\  
NGC 3850 & 176.39817 & 55.88689 & 0.00386 & 109  & 485  & 0.03 \\  
SDSS J115153.66+530558.2 & 177.97360 & 53.09951 & 0.00359 & 29  & 496  & 0.004 \\  
MESSIER 109 & 179.39992 & 53.37453 & 0.00348 & 0  & 506  & 2.0 \\  
SDSS J120139.61+551231.0 & 180.41508 & 55.20864 & 0.00398 & 146  & 513  & 0.006 \\  
2MASX J12015013+5508422 & 180.45881 & 55.14491 & 0.00370 & 61  & 515  & 0.01 \\  
UGC 06988 & 179.96546 & 55.66533 & 0.00240 & -325  & 515  & 0.03 \\  
LBT J115205.6+544732.2 & 178.02326 & 54.79229 & 0.00107 & -725  & 23  & - \\  
\enddata

\tablenotetext{a}{Columns 6 and 7 contain the impact parameter and fraction of L$_*$, respectively, calculated at a distance of 18.6 Mpc.  The exception is the Grapes, which is calculated at the luminosity distance implied by a Hubble flow with recession velocity corrected per \citet{Mould:2000vn}.}
\end{deluxetable*}

\end{document}